\documentclass[prl,showpacs,preprintnumbers,amsmath,amssymb,tightenlines,twocolumn]{revtex4}

\usepackage{graphicx}
\usepackage{dcolumn}
\usepackage{bm}
\usepackage{epsfig}
\usepackage{stmaryrd}

\newcommand{\ssection}[1]{{\noi  \it #1:}}
\newcommand{\bra}[1]{\langle\,{#1}\, |}
\newcommand{\ket}[1]{|\,{#1}\,\rangle}

\newcommand{\pdiff}[2]{\frac{\partial #1}{\partial #2}}
\newcommand{\pdiffn}[3]{\frac{\partial^{#3} #1}{\partial #2^{#3}}}

\newcommand{\sub}[2]{{#1}_{\mbox{\!\! \scriptsize #2}}}

\newcommand{\bv}[1]{\mathbf{ #1 }}

\def\noi{\noindent}
\def\beq{\begin{equation}}
\def\eeq{\end{equation}}

\def\CR{\nonumber\\[0.15cm]}

\newcommand{\fref}[1]{Fig.~\ref{#1}}
\newcommand{\frefp}[2]{Fig.~\ref{#1}~(#2)}

\newcommand{\eref}[1]{Eq.~(\ref{#1})}

\newcommand{\cref}[1]{chapter~\ref{#1}}

\newcommand{\Cref}[1]{Chapter~\ref{#1}}

\newcommand{\bref}[1]{(\ref{#1})}

\usepackage{ulem}  
\usepackage{color}

\begin{document}

\title{Conical intersections in an ultracold gas}
\author{S.~W\"uster}
\affiliation{Max Planck Institute for the Physics of Complex Systems, N\"othnitzer Strasse 38, 01187 Dresden, Germany}
\author{A.~Eisfeld}
\affiliation{Max Planck Institute for the Physics of Complex Systems, N\"othnitzer Strasse 38, 01187 Dresden, Germany}
\author{J.~M.~Rost}
\affiliation{Max Planck Institute for the Physics of Complex Systems, N\"othnitzer Strasse 38, 01187 Dresden, Germany}
\email{sew654@pks.mpg.de}

\begin{abstract}
We find that energy surfaces of more than two atoms or molecules interacting via dipole-dipole potentials generically possess conical intersections (CIs). Typically only few atoms participate strongly in such an intersection. For the fundamental case, a circular trimer, we show how the CI affects adiabatic excitation transport via electronic decoherence or geometric phase interference. These phenomena may be experimentally accessible if the trimer is realized by light alkali atoms in a ring trap, whose dipole-dipole interactions are induced by off-resonant dressing with Rydberg states. Such a setup promises a direct probe of the full many-body density dynamics near a conical intersection.
\end{abstract}


\maketitle

\ssection{Introduction}
Conical intersections (CIs) of electronic energy surfaces are a generic feature of large molecules \cite{yarkony2001conical, worth:CI:review}. The intersections provide fast \emph{intra-molecular} transition channels between electronic states and can thus affect the outcome of (photo-) chemical processes. Radiation-less de-excitation of large bio-molecules proceeds through these channels and yields enhanced photostability that might have been crucial for the development of life on earth \cite{applegate2003explorations,Ismail-Robb-Ultrafastdecayof-2002,Perun-Domcke-Abinitiostudies-2005}. Modern techniques allow the theoretical investigation of quantum wave-packet dynamics at conical intersections of large molecules in quite some detail \cite{burghardt:MCTDH_pyrazine,Worth-Burghardt-UsingMCTDHwavepacket-2008}. However experiments usually monitor such dynamics \emph{indirectly}, e.g., through reactant fractions or fluorescence spectra. 

The \emph{direct} observation of many-body densities could be realized near conical intersections in ultracold gases. As we demonstrate, these are ubiquitous between energy surfaces of \emph{inter-atomic} (or molecular) dipole-dipole interactions \cite{FiSeEn08_12858_}. Such interactions are responsible for excitation transport in photosynthetic light harvesting units \cite{grondelle:review}, molecular aggregates \cite{eisfeld:Jagg} or cold Rydberg gases \cite{anderson:resonant_dipole}. We find that typical ensembles have sub-units of three to seven particles that are mainly responsible for the CI. We investigate the fundamental unit, a ring-trimer, in more detail. In the ring geometry, two adiabatic energy surfaces cross at the equilateral triangle configuration, forming a conical intersection. While a direct crossing of the CI splits wave packets among two electronic surfaces and thus de-coheres the reduced electronic state, circumnavigation of the CI results in interference patterns with clear signatures of the geometric phase. 

We suggest a realization of the ring-trimer using trapped atoms that obtain dipole-dipole interactions through off-resonant Rydberg dressing \cite{wuester:dressing}. Through spatially resolved detection of single atoms one can reveal the full many-body dynamics around the CI.
\begin{figure}[htb]
\centering
\epsfig{file={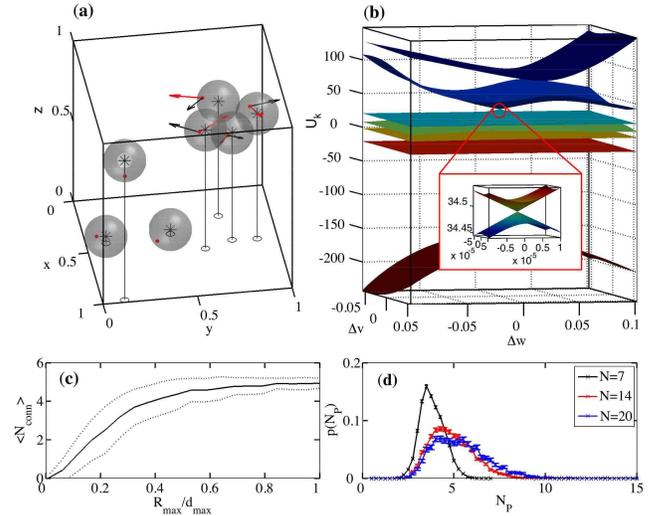},width=0.99\columnwidth} 
\caption{(color online) Conical intersections in ensembles of dipole-dipole interacting atoms or molecules. (a) Randomly placed particles
 ($\ast$) and location of a selected conical intersection (red $\bullet$) in their vicinity (grey shade, $R_{max}=0.2d_{max}$, see text). (b) Potential energy surfaces $U_{k}(\Delta v\bv{V} + \Delta w \bv{W})$ near the CI in (a) using $\mu=1$. The modes $\bv{V}$  ($\bv{W}$) shown as red (black) arrows in (a) span the CI's branching plane, see \eref{branching_plane}. (c) Mean connectivity $\sub{N}{conn}$ (solid line), as explained in the text, and its standard deviation (dotted lines). (d) Histogram of participation numbers for $\sub{R}{max} = 0.35\sub{d}{max}$.
\label{generic_CIs}}
\end{figure}

\ssection{Conical intersections in dipole-dipole gases}
Consider a sample of $N$ atoms or molecules of mass $M$ whose positions $\bv{r}_{n}$ are grouped into the $3N$ component vector $\bv{R}=\{\bv{r}_n\}$. We assume that each particle can be either in some electronic ground state $\ket{g}$ or some excited state $\ket{h}$. The single excitation Hilbert space for the electronic degree of freedom is then spanned by $\ket{\pi_n}\equiv \ket{g\cdots h\cdots g}$, with the $n$'th particle excited. We incorporate particle motion and electronic excitation transfer by dipole-dipole interactions \cite{noordam:interactions} into the Hamiltonian
\begin{align}
\hat{H}&=-\sum_n^N \frac{ \nabla_{\bv{r}_{n}}^2}{2 M}  +  \hat{H}_{el}(\bv{R}),
\label{AggregateHamiltonian}
\end{align}
with
\begin{align}
\hat{H}_{\rm el}(\bv{R})&=\sum_{n,m}^N  V_{nm}(\bv{r}_{n},\bv{r}_{m}) | \pi_{n}\rangle \langle \pi_{m}|,
\label{ElectronicHamiltonian}
\end{align}
where $V_{nm}$ is the dipole-dipole interaction between the particles. Besides Rydberg atoms \cite{cenap:motion,wuester:cradle}, this model can also describe molecular aggregates \cite{eisfeld:Jagg} or Rydberg-dressed ground state atoms \cite{wuester:dressing}. We ignore the angular dependence of the interaction and specify $V_{nm}=-\mu^2/|R_{nm}|^3$ with $R_{nm}=|\bv{r}_{n}-\bv{r}_{m}|$.  Atomic units are implied unless otherwise indicated.

To analyze the Hamiltonian \bref{AggregateHamiltonian}, consider a Born-Oppenheimer separation of dynamics \cite{domke:yarkony:koeppel:CIs}. The eigenstates of the
electronic Hamiltonian fulfill $H^{\rm el}(\bv{R})\ket{\varphi_{j}(\bv{R})}=U_{j}(\bv{R} )\ket{\varphi_{j}(\bv{R})}$. The energies $U_{j}(\bv{R})$ then define
$N$ adiabatic potential energy surfaces for particle motion. A particle stays on a definite surface, if non-adiabatic couplings between the surfaces can be ignored. These non-adiabatic couplings become large at a conical intersection.
 
To assess the relevance of conical intersections in an ultracold gas, we consider an ensemble of $N$ atoms distributed randomly in a cube of
side length $L$, with coordinates $\bv{R_*}$ and an isotropic quantum or thermal position uncertainty of about $\sigma_{R^*}$.
After determination of $\bv{R_*}$, we seek CI-locations $\bv{R}_{CI}$ within a distance
 $R_{max}=\sigma_{R^*}$\cite{footnote:Rmaxdef} by numerical minimization of the smallest separation between any
 two surfaces $U_{i}$ and $U_{j}$.  We define $\sub{N}{conn}$ as the number of adjacent energy
surfaces between which we find a CI under these constraints. Its average is shown in \frefp{generic_CIs}{c} as a function of $R_{max}/d_{max}$,
where $d_{max} =L/(2 \pi N)^{1/3}$ is the peak of the nearest neighbor distance distribution. We averaged over $50$ random atomic ensembles and
$200$ different minimization seeds. Already for $R_{max}
\sim 0.2 d_{max} \sim 0.06L$ at least $3$ CIs are accessible on average. An ensemble with spread of position $\sigma_{R^*}\sim 0.2 d_{max}$, would necessarily access these CIs.

To characterize CIs between surfaces $i$ and $j$ in more detail, consider their branching plane spanned by~\cite{yarkony2001conical}
\begin{align}
2 \bv{v}^{ij}_{n} &= \bra{\varphi_{i}}\bv{\nabla}_{\bv{r}_n} \hat{H}_{el}\ket{\varphi_{i}} + \bra{\varphi_{j}}\bv{\nabla}_{\bv{r}_n} \hat{H}_{el}  \ket{\varphi_{j}},
\CR
\bv{w}^{ij}_{n} &= \bra{\varphi_{i}}\bv{\nabla}_{\bv{r}_n} \hat{H}_{el}\ket{\varphi_{j}}.
\label{branching_plane}
\end{align}
Let $\bv{V}^{ij}=\{\bv{v}^{ij}_n\}$, $\bv{W}^{ij}=\{\bv{w}^{ij}_n\}$ be $3N$ component vectors and $\bar{\bv{V}}^{ij} =\bv{V}^{ij}/|\bv{V}^{ij}|$, $\bar{\bv{W}}^{ij} =\bv{W}^{ij}/|\bv{W}^{ij}|$. 
The motion of atom $m$ has a large effect on the energy gap between surfaces $i$ and $j$, if the $|\bar{\bv{v}}_m|^2$ or $|\bar{\bv{w}}_m|^2$ components of the branching vectors are large.
Motion in the \emph{seam-space} orthogonal to $\bar{\bv{V}}^{ij}$, $\bar{\bv{W}}^{ij}$ has almost no effect, thus small $|\bar{\bv{v}}_m|^2$ or $|\bar{\bv{w}}_m|^2$ indicate that the atom does not participate strongly in the intersection.
This motivates the use of a participation number (see e.g.~\cite{vlaming:disorder}) $N_{P}^{ij}=(1/\sum_{n}|\bar{\bv{v}}^{ij}_{n}|^4  + 1/\sum_{n}|\bar{\bv{w}}^{ij}_{n}|^4)/2$, which roughly gives the number of atoms involved in CI dynamics. The $N_{P}$-histogram in \frefp{generic_CIs}{d} shows that these are usually three to seven. Changes in density $\rho=N/L^3$ only affect the overall energy scale of \frefp{generic_CIs}{b} but not the results in \frefp{generic_CIs}{c-d} if distances are expressed in units of $\sub{d}{max}$.

We conclude that conical intersections are generic in dipole-dipole interacting aggregates with moving constituents. This holds also for tightly bounded motion, such as small vibrations around fixed locations. Since a small number of atoms control the CIs we will study the fundamental case $N=3$ in the following.

\begin{figure}[htb]
\centering
\epsfig{file={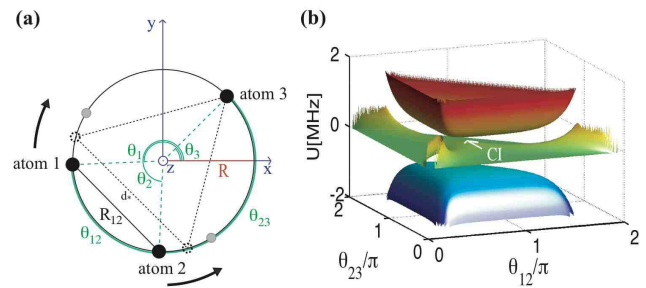},width=0.99\columnwidth} 
\caption{(color online) Ring trimer. (a) Three atoms confined to a ring with radius $R$ in the $x$-$y$ plane. (b) Born-Oppenheimer surfaces for the electronic Hamiltonian \bref{AggregateHamiltonian}. From top to bottom $\sub{U}{rep}$, $\sub{U}{mid}$ $\sub{U}{att}$. For better visibility we shifted the repulsive (attractive) surface up (down) by $\Delta U=0.37$ MHz, and magnified the middle surface by a factor $f_{m}=5$. At the marked CI location, $\sub{U}{rep}$ and $\sub{U}{mid}$ would touch without the shifts.
\label{system_sketch}}
\end{figure}

\ssection{Ring trimer}
We consider three particles confined one-dimensionally on a ring, with coordinates as in \fref{system_sketch}. On the ring the position of atom $n$ is specified by its 2D polar angle $\theta_{n}$. Then the dipole-dipole interaction is 
$
V_{nm}=- \mu^{2}/(R \sqrt{2(1-\cos[\theta_{n} - \theta_{m}])} )^3.
$
The centre-of-mass type angle $\theta_{CM}=\sum_{n} \theta_{n}/3$ decouples, and the dynamics of interest is fully described by the two relative angles $\theta_{12}=\theta_{2}-\theta_{1}$ and $\theta_{23}=\theta_{3}-\theta_{2}$. 

In \fref{system_sketch} we show the three potential energy surfaces obtained by diagonalizing $\hat{H}_{\rm el}$. 
The top (red) and bottom (blue) surfaces are globally repulsive or attractive, respectively. Each of them is associated with a different delocalized excitation \cite{cenap:motion}. Energetically between these, we find a middle surface that touches the repulsive one in a conical intersection at $\theta_{12}=\theta_{23}=2\pi/3$. 
In the following we show how this intersection affects adiabatic excitation transport. 

The total quantum state is $\ket{\Psi(\bv{R})}=\sum_{n=1}^{N} \phi_{n}(\bv{R}) \ket{\pi_n}$, where $\phi_{n}(\bv{R})$ is the wave function of particle motion in the electronic state $ \ket{\pi_n}$. 
From the Hamiltonian \bref{AggregateHamiltonian}  one obtains the equations
\begin{align}
\label{fullSE}
i\frac{\partial}{\partial t}   \phi_{n} (\bv{R})&=\sum_{m=1}^N\left[-\frac{ \nabla^2_{R_m} }{2 M}  \phi_{n}(\bv{R}) + V_{nm}(R_{nm}) \phi_{m}(\bv{R})\right].
\end{align}
For $N=3$ atoms we use $\theta_{12}$ and $\theta_{23}$ and write $\ket{\Psi}=\sum_{n=1}^{3} \phi_{n}(\theta_{12},\theta_{23})\ket{\pi_{n}}$, which reduced \eref{fullSE} to 
\begin{align}
&i \frac{\partial}{\partial t}  \phi_{n}=\left[-\pdiffn{}{\theta_{12}}{2} + \pdiff{}{\theta_{12}} \pdiff{}{\theta_{23}}   - \pdiffn{}{\theta_{23}}{2}  \right] \phi_{n}
\CR
&- \sum_{m=1}^{3} V_{nm}(\theta_{12},\theta_{23})\phi_{m}.
\label{threeatom_TwoDGF_SE}
\end{align}
We also introduce the adiabatic form of the wavefunction $\ket{\Psi}=\sum_{j=1}^{3} \tilde{\phi}_{j}(\theta_{12},\theta_{23}) \ket{\varphi_{j}(\theta_{12},\theta_{23})}$. Finally, let us define the population of the $n$'th diabatic (adiabatic) state as $p_{n}=\int d\bv{R} |\phi_{n}|^{2}$ ($\tilde{p}_{j}=\int d\bv{R} | \tilde\phi_{j}|^{2}$). For adiabatic states and surfaces, we will use subscripts $\mbox{rep}=1$, $\mbox{mid}=2$, $\mbox{att}=3$ as in \fref{system_sketch}. The setup described above allows one to observe how CIs affect many-body dynamics. We will discuss two paradigmatic situations.

\ssection{Electronic de-coherence}
\label{CI_dynamics}
A robust consequence of the CI is to decohere reduced electronic state of the the trimer. Consider the case where initially the system is on the repulsive adiabatic surface and the particles form an isosceles triangle as skeched in \frefp{system_sketch}{a}. The initial distance between particles $1$ and $2$ is $R_{12}=d_{0}<d^{*}$, where $d^{*}=\sqrt{3}R$ is the separation at the conical intersection. Each atom has initially a Gaussian angular distribution with width $\sigma_{0}\ll 2\pi$. The detailed construction of the initial state is described in \cite{wuester:cradle}. From this initial state, the repulsion drives the configuration over the CI. We use $R=9.8\mu$m, $\mu=180$ a.u.~and $M=11000$. These parameters correspond to laser dressed Lithium, as will be explained later.

\begin{figure}[htb]
\centering
\epsfig{file={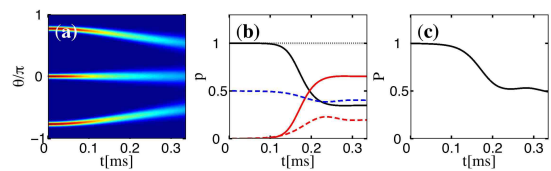},width=0.99\columnwidth} 
\caption{(color online) (a) Total atomic density, given by $n(\theta)= \sum_{m=1}^N n_{\{ m \} }(\theta_{m})|_{\theta_{m} \rightarrow \theta}$ with $n_{\{ m \} }(\theta_{m}) = \sum_{n=1}^N \int d^{N-1}\theta_{\{m\}} |\phi_{n}(\bv{\theta})|^{2}$, where $\int d^{N-1}\theta_{\{m\}}$ denotes integration over all but the $m$'th particle coordinate. Note that the $\theta$-axis is periodic, hence the highest and lowest particle are near neighbors and initially repel. The configuration passes the CI location when all three inter-particle distances are equal, around $t=0.17$ ms. (b) Adiabatic populations $\sub{p}{rep}$ (solid, black), $\sub{p}{mid}$ (solid, red), electronic populations $\sub{p}{1}=\sub{p}{3}$ (dashed, blue), $\sub{p}{2}$ (dashed, red) and total population (dotted, black). (c) Purity $P=\mbox{Tr}[\hat{\sigma}^2]$ of the reduced electronic density matrix, described in the text. 
\label{numbers_crossing}}
\end{figure}

The close proximity pair of particles initially accelerates adiabatically, as evident from the constant surface populations in \frefp{numbers_crossing}{b}. However, as the many-body wave function passes the conical intersection, more than $60\%$ of the total population is transferred from the repulsive to the middle surface. \frefp{numbers_crossing}{c} shows an accompanying drop in purity of the reduced electronic density matrix $\hat{\sigma}=\sum_{n,m}\sigma_{nm} \ket{\pi_{n}}\bra{\pi_{m}}$, with $\sigma_{nm}=\int d^N \bv{R}\:\:  \phi^*_{n}(\bv{R}) \phi_{m}(\bv{R})$. The operator $\hat{\sigma}$ describes the electronic state, disregarding knowledge of particle positions. The observed decoherence of the electronic state is a consequence of the spatially disjunct splitting of the wave-packet on two surfaces and ensuing entanglement between the particle position and aggregate electronic state. It will occur generically for all dynamics close to the CI.
%

\begin{figure}[htb]
\centering
\epsfig{file={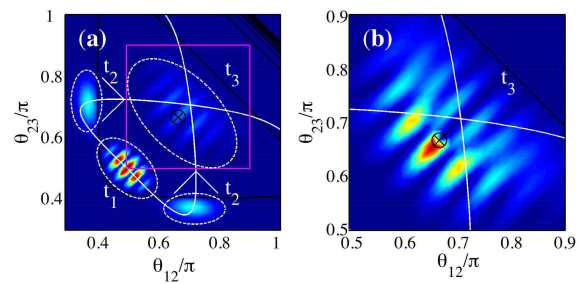},width=0.99\columnwidth} 
\caption{(color online) Intersection circumnavigation from initial momentum superposition. The resulting interference pattern manifests the geometrical phase. (a) Shown is the density $|\sub{\tilde{\phi}}{rep}|^2$ at times $t_{1}=0$, $t_{2}=0.24$ms, $t_{3}=0.48$ms. The white lines are classical trajectories on the repulsive surface. Segregation of different time samples is provided by the white dashed ellipses. (b) Magnification of the area within the violet square in (a). A $\pi$-shift of the pattern due to the geometrical phase is visible between the part of the wave packet that has circumnavigated the CI ($\varotimes$) and the part that has not. Black lines in the background are iso-contours of $U_{rep}$.
\label{berry_run}}
\end{figure}

\ssection{Geometrical phase}
A more subtle but even more fundamental consequence of conical intersections is the geometric (Longuet-Higgins-Berry) phase picked up by wave-packets that are adiabatically transported in a closed-loop around them \cite{berry:phase,longuet:higgins:phase}. In our case, wave-packets moving half a circle around the CI on the repulsive surface with opposite sense of rotation meet with a relative phase of $\pi$. We initially superimpose two different momentum components, which yield the classical trajectories shown in \frefp{berry_run}{a} as solid white lines. The resulting quantum interference pattern after encircling the CI is shown in \frefp{berry_run}{b} for $R=13\mu$m.
The wave-packet is chosen to contain components that will \emph{not} actually have enclosed the CI upon interference. These form the lower left part of the pattern in \frefp{berry_run}{b}, with an antinode on the diagonal. In contrast, the top right of the pattern resulting from motion enclosing the CI has a node on the diagonal. This manifestation of Berry's phase shift can be experimentally accessed due to the control and detection afforded by ultracold atoms as we will show in the next section. The interference pattern could then be used as a sensitive probe of the many-body energy landscape, as for example the presence of a fourth perturber atom would move the CI out of the $\theta_{12}$,  $\theta_{23}$ plane and hence yield geometric phases differing from $\pi$ \cite{yarkony2001conical}. 

\ssection{Laboratory realisation}
We propose a realisation of the ring-aggregate model using trapped alkali atoms. To increase flexibility in the choice of parameters, we assume that long-range interactions are induced by dressing with Rydberg states \cite{wuester:dressing}, which allows the atoms to be predominantly in their ground-state. For this we require two non-interacting ground (or meta-stable) states $\ket{g}$, $\ket{h}$, each of which is off-resonantly coupled to a Rydberg level $\ket{g}\leftrightarrow \ket{s}$, $\ket{h}\leftrightarrow \ket{p}$ by two separate transitions, both with Rabi-frequency $\Omega$ and detuning $\Delta$. The detuning parameter $\alpha= \Omega/2 \Delta $ controls the population transferred to the Rydberg levels $\ket{s}$, $\ket{p}$, of order $\sim \alpha^2$, and the interaction-strength for the ground state atoms ($\sim \alpha^{4}$). 

Let the Rydberg levels $\ket{s}$, $\ket{p}$ have a principal quantum number $\nu$. We parameterize the dipole-dipole interaction strength $\mu$ and life-time $\tau$ of the dressed Rydberg states as $\mu=\mu_{0}\nu^{2}\alpha^{2} $ and $\tau = \tau_{0} \nu^{3}/\alpha^2$ respectively, where $\mu_{0}$ and $\tau_{0}$ can be found from a reference Rydberg level. We will explicitly consider $^{7}$Li with $M=11000$, $\mu_{0}=0.8$, $\tau_{0}=3\times 10^{7}$ \cite{theodosiou:Li:life} (all in atomic units).

The simulations presented above correspond to $^{7}$Li atoms confined on a ring of radius $R=9.8\mu$m, acquiring long-range interactions through laser dressing with $\alpha=0.15$ via the Rydberg level $\nu=100$.
The parameters have been chosen to meet the following conditions:
\\ \noi
{\it 1. The dipole-dipole interaction energy does not exceed realistic transverse oscillator spacings $\omega_{\perp}$ of the ring, enabling one-dimensional dynamics.} The effects presented can occur close to the CI where the 
 atoms are separated by $d_{0}\approx  d_{*}$, their separation at the CI. Their dressed interaction energy is then $V(d_{*})=\mu^{2} / d_{*}^{3} = \mu_{0}^{2}  \nu^{4}\alpha^{4}  /( \sqrt{3}R)^{3}=40$ kHz. This is within reach of very tight trapping potentials. 
\\ \noi
{\it 2. The \emph{bare} dipole-dipole shift without dressing is smaller than the dressing laser detuning, at the closest approach of the atoms.} Otherwise resonances would yield undesirably large excited state populations. We have $\sub{V}{bare}(d_{*})=V(d_{*})/\alpha^4=80$MHz, while a detuning of $\Delta \approx 150$ is feasible for $\nu=100$ \cite{wuester:dressing}. 
\\ \noi
{\it 3. The motion is adiabatic, except close to the conical intersection,} as seen in \fref{numbers_crossing}.
\\ \noi
{\it 4. The life-time of the dressed three-atom system ($\tau = \tau_{0} \nu^{3}/\alpha^2 /N =10$ms) exceeds the duration of the motion.}  
\\ \noi
A systematic search for regions in parameter space where realistic ring-trimers exist is non-trivial and will be reported elsewhere. It shows that light alkali atoms such as Li are favorable and that dressing is crucial. Without it, the above conditions can only be fulfilled for unrealistically large principal quantum numbers. 

In conclusion, we have shown that \emph{external} conical intersections in dipole-dipole interacting gases of ultracold (Rydberg) atoms or molecules are common, as in the \emph{internal} energy landscape of large molecules. We have presented a setup which holds promise for direct laboratory studies of the many-body wave function dynamics near or across a conical intersection. This can be achieved in a circular aggregate of Rydberg-dressed alkali atoms confined in a 1D ring geometry, for which we have worked out paradigmatic consequences of the CI such as electronic decoherence and nuclear interference due to the geometric phase. Conical intersections in laboratory co-ordinates can also be designed through external fields for ultra-cold molecules \cite{wallis2009conical,moiseyev:CI_lattice}, but lack by construction the many-body interacting character of the CIs studied here. 

Similar effects as in our examples would occur also if the dipole-dipole interacting particles were not free on a ring, but confined in separate potential wells, as long as vibrations within each well allow many-body wave packet excursions over the CI. Hence, our results pertain in particular to molecular aggregates such as light harvesting units.

\end{document}